\documentclass[%
 reprint, 
 amsmath, amssymb, 
 aps, 
]{revtex4-2}

\usepackage{graphicx}
\usepackage[citecolor=blue,colorlinks=true,linkcolor=blue]{hyperref}
\usepackage{dcolumn}
\usepackage{bm}
\usepackage{hyperref}
\usepackage{braket} 
\usepackage[mathlines]{lineno}

\begin{document}

\preprint{APS/123-QED}

\title{Magnetic Skyrmion Encoding by Structured Light}

\author{Qifan Zhang\(^{1,2,3}\)}
\author{Wangke Yu\(^{4}\)}
\author{Zhongquan Nie\(^{6}\)}
\author{Yijie Shen\(^{4,5,}\)}
\email{yijie.shen@ntu.edu.sg}
\author{Shirong Lin\(^{1,2,}\)}
\email{shironglin@gbu.edu.cn}

\affiliation{\makebox[0pt][l]{\textsuperscript{1}} School of Physical Sciences, Great Bay University, Dongguan 523000, China}%

\affiliation{\makebox[0pt][l]{\textsuperscript{2 }} Great Bay Institute for Advanced Study, Dongguan 523000, China}

\affiliation{\makebox[0pt][l]{\textsuperscript{3 }} School of Physics and Materials Science, Guangzhou University, Guangzhou 510006, China}%

\affiliation{\makebox[0pt][l]{\textsuperscript{4}} Centre for Disruptive Photonic Technologies, School of Physical and Mathematical Sciences, Nanyang Techno logical University, Singapore 637371, Singapore}

\affiliation{\makebox[0pt][l]{\textsuperscript{5}} School of Electrical and Electronic Engineering, Nanyang Technological University, Singapore 639798, Singapore}

\affiliation{\makebox[0pt][l]{\textsuperscript{6}} National University of Defense Technology, College of Advanced Interdisciplinary Studies, Changsha 410073, China}






\begin{abstract}
Structured light fields, featuring unique topological properties and high tunability, have opened new frontiers in light-matter interactions with magnetic systems. However, the ultrafast and reconfigurable optical encoding of various types of topological magnetic textures remains a significant challenge. Here, we systematically investigate the encoding mechanism of structured light in magnets via the higher-order Poincaré sphere. By uncovering the precise relationship between the winding number of structured light and the topological charge of magnetic textures, we establish a fundamental topological connection between light and magnetism. This framework enables ultrafast, all-optical encoding of diverse topological spin textures in magnetic media, including skyrmions, antiskyrmions and skyrmion bags. Our work advances the fundamental understanding and all-optical control of topological magnetism, offering a promising route for designing skyrmion-based devices.
\\

\end{abstract}

\maketitle

\(Introduction\)—Structured light \cite{forbes2021structured,he2022towards,shen2019optical,milione2011higher,zhan2006properties,ghosh2023ultrafast,truc2023light,tang2020harmonic,lin2022all,chen2021engineering,chen2024highly,mei2023cascaded,miao2023deterministic,wang2024topological}, owing to its diverse topological properties and high degree of controllability, holds immense potential not only for fundamental studies but also in the field of light-controlled magnetism \cite{lin2019all,fanciulli2025magnetic,kim2025, finazzi2013laser,guang2020creating,tm58-lbdl,nukui2025light,guo2024single,wang2021dual,zhang2025deterministic,wu2023topology,assouline2025faraday,PhysRevLett.117.147202,assouline2024helicity,zhou2025contrasting}, particularly in the manipulation of magnetic topological textures such as magnetic skyrmions \cite{leonov2016properties,fujita2017encoding,zhang2026,mallick2024oam,bhowal2022magnetoelectric,wang2020plasmonic,zhu2024ultrafast,gerlinger2021application,PhysRevB.107.144429,miyata2022topological,shi2024dynamic,hu2025recent}. Magnetic skyrmions are topological spin textures in real space \cite{jiang2017direct,schulz2012emergent,wu2020ferrimagnetic,yokoyama2015josephson,mascot2021topological,PhysRevB.111.115115,dohi2025observation,zheng2023hopfion,jin2024skyrmion}, and are characterized by their small size \cite{wang2021stripe,wang2018theory,wu2021size,wang2023topological}, versatile generation \cite{amin2023antiferromagnetic,buttner2021observation,koshibae2014creation,yokouchi2020creation,khela2023laser,titze2024all,li2024room,juge2022skyrmions,samardak2018enhanced,ding2025multistep,bhattacharya2020creation,li2026,chen2025magnetic,kalin2024optical} and manipulation methods \cite{li2024electron,jin2024rolling,bernstein2025spin,kuchkin2025symmetry,brems2025realizing,liu2025dynamics,dou2025ultrafast,cai2024acceleration,mackinnon2022collective,shen2020current,niu2025magnetic,yu2023voltage,li2020tunable}. Additionally, skyrmions exhibit fluidity, enabling multiple skyrmions to form skyrmion bags \cite{rybakov2019chiral,foster2019two,wang2023particle,wu2022nematic,yang2024embedded,jiang2025stable,liu2025room}. In the field of information storage, skyrmions demonstrate significant application potential \cite{fert2013skyrmions,dai2026transient,ma2025domain,chang2025suppressed,petrovic2025colloquium,back20202020}.

In 2017, it was shown that circularly polarized vortex light can controllably generate magnetic skyrmions in chiral magnets \cite{fujita2017encoding}. 
Recently, our group revealed the mechanism whereby vortex light generates a specific number of magnetic skyrmions, determined by the angular momentum and amplitude of light \cite{zhang2026}.
Yet, research on generating skyrmions with structured light needs expansion in breadth (to diverse polarizations and materials) and depth (mechanism). A key limitation is the lack of a defined angular momentum for beams with arbitrary polarization patterns. To our knowledge, a significant gap exists in both applying such structured
light to encode skyrmions and in understanding the corresponding mechanisms. In 2011, Milione et al. proposed the higher-order Poincaré sphere (HOPS) to describe the higher-order polarization states of vector vortex beams \cite{milione2011higher}. The model provides a concise, unified framework for structured light-matter interactions, offering new perspectives for light-controlled magnetism.

Here, we investigate the encoding of magnetic topological textures in ferromagnetic thin films using topological HOPS structured light. We develop a light-magnet coupling model in which the structured light’s magnetic field is incorporated via the Zeeman effect (Our model also applies to light's electric field in multiferroics via magneto-electric coupling \cite{fujita2017encoding}), and simulate magnetization dynamics with the Landau–Lifshitz–Gilbert (LLG) equation. The light’s magnetic field exhibits topological features quantified by a winding number. Through interaction with the magnetic medium, controlled encoding of skyrmions, antiskyrmions, and skyrmion bags is achieved in magnet with different Dzyaloshinskii–Moriya interaction (DMI) types. The topology of encoded textures correlates with the winding number, highlighting implications for light-controlled skyrmionics.
\begin{figure*}
\centering
\includegraphics[width=1\linewidth]{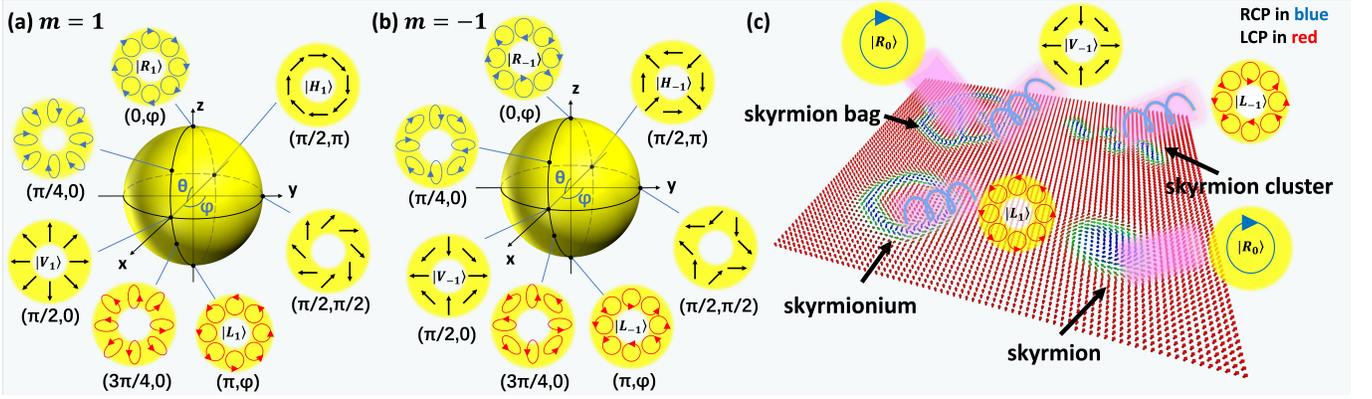}
\caption{\label{Fig1}Representation of the higher-order Poincaré sphere (HOPS) for (a) \(m=1\) and (b) \(m=-1\). Representative polarization states at selected (\(\theta\), \(\varphi\)) points are illustrated. The poles correspond to circularly polarized vortex eigenstates, the equator corresponds to \(\pi\)-vector beams, and intermediate points correspond to elliptically polarized states. (c) Schematic illustration of magnetic textures encoded by HOPS structured light. 
}
\end{figure*}

\(Model\)—The north and south poles of the HOPS represent two orthogonal circularly polarized vortex eigenstates carrying opposite topological charges, which constitute the fundamental components of any state of polarization in the sphere. The amplitude distribution of the magnetic fields of structured beams on the poles of HOPS can be expressed as
\begin{equation}
B(\bm{\mathrm{r}}, t) = B_0 \dfrac{A_{m}(\bm{\mathrm{r}})}{\max_{\bm{\mathrm{r}}} \left|A_{m}(\bm{\mathrm{r}}) \right|}e^{({-{{\frac{(t-t_0)^2}{\sigma^2}}}-{i\omega t}})} . \tag{1}
\end{equation}
Here, \(A_{m}\) is the amplitude distribution of the magnetic field generated by the Laguerre–Gaussian beam, \(m\) the value of orbital angular momentum (OAM) and also the order of the HOPS, \(B_0\) the magnetic field strength coefficient, \(\omega\) the light frequency, \(t_0\) the time at which the magnetic field strength reaches its maximum within one period, and \(\sigma\) the pulse duration. We work in the paraxial regime and consider only the transverse magnetic-field components \(\bm{\mathrm{B}}=B(\bm{\mathrm{r}}, t)\mathrm{\bm{e}_p}\), where \(\mathrm{\bm{e}_p}=\bm{\mathrm{\hat{x}}}\pm i\bm{\mathrm{\hat{y}}}\). Here we construct a time-localized pulse by multiplying the monochromatic HOPS field at a carrier frequency (\(\omega\)) by a Gaussian temporal envelope \cite{fujita2017encoding,zhang2026}. 

In the following analysis, we let \(\theta\) and \(\varphi\) denote the latitude and longitude angles on the HOPS, \(\Theta\) the azimuth angle in the beam cross section and \(m\) the topological charge, which also indicates the order of the HOPS.
Any point on the sphere corresponds to a polarization state that can be expressed as a linear superposition of the two eigenstates. This relationship can be described by the following two-dimensional Jones vector \cite{milione2011higher}: 
\begin{equation}
\ket{\bm{\mathrm{\rho}}}=\mathrm{cos}(\frac{\theta}{2})e^{-i\varphi}\ket{R_{m}}+\mathrm{sin}(\frac{\theta}{2})\ket{L_{m}},\tag{2}
\end{equation}
with respect to the orthonormal circular polarization basis
\(\{R_{m}, L_{m}\}\) such that \(\ket{R_{m}}=\mathrm{exp}(-im\Theta)(\bm{\mathrm{\hat{x}}}+i\bm{\mathrm{\hat{y}}})\) and \(
\ket{L_{m}}=\mathrm{exp}(im\Theta)(\bm{\mathrm{\hat{x}}}-i\bm{\mathrm{\hat{y}}})\). From Equations (2), we obtain the expression for the magnetic field distribution from a HOPS structured beam with order \(m\) as follows:
\begin{equation}
\begin{aligned}
\ket{\bm{\mathrm{B}}_m}=\mathrm{cos}(\frac{\theta}{2})[e^{-i(m\Theta+\varphi)},ie^{-i(m\Theta+\varphi)}]^{T}\\+\mathrm{sin}(\frac{\theta}{2})[e^{im\Theta},-ie^{im\Theta}]^{T}.
\end{aligned} \tag{3}
\end{equation}
According to Equation (3), we observe that each term contains \(m\Theta\). This indicates that, for the HOPS structured beam with order \(m\), along a closed path around the optical vortex in the transverse plane (\(\Theta\) from \(0\) to \(2\pi\)), the beam’s complex amplitude accumulates a total phase \(\Phi\), which is \(2\pi\cdot m\).
The topological properties of the magnetic field distributions from the HOPS structured beam with order \(m\) can be characterized using a winding number, which is calculated by the following formula \cite{zhan2006properties}: 
\begin{equation}
l=\frac{1}{2\pi}{\oint\limits}_C\nabla\Phi \cdot ds,\tag{4}
\end{equation}
where \(C\) is a circular closed loop enclosing the vortex center, and \(ds\) is the path element along \(C\). From Fig. \ref{Fig1} (a) and (b), as the azimuthal angle \(\Theta\) varies from \(0\) to \(2\pi\) around the center, the instantaneous magnetic polarization direction rotates accordingly. The number and direction (counterclockwise/clockwise) of rotations determine the absolute value and sign (\(+\)/\(-\)) of the winding number. Therefore, the magnetic field distributions generated by all light beams on the same HOPS structured beam with order \(m\) share identical topological properties. 

Here, we focus on the HOPSs for \(m=1\) and \(m=-1\), as shown in Fig. \ref{Fig1} (a) and (b), respectively. The plots display the magnetic field distributions of light corresponding to several representative points on the spheres. 

For example, when \(\theta=\pi/2\), it corresponds to \(\pi\)-vector beams on the equator:
When \(m=1\) and \(\varphi=0\), according to Equations (3) and (4), \(\ket{\bm{\mathrm{B}}_1}=[\mathrm{cos}\Theta, \mathrm{sin}\Theta]^T\) and \(l=1\). In this case, the magnetic field distribution remains radial and appears as a single continuous annular domain without polarization segmentation. 
When \(\varphi\neq0\), the magnetic field direction remains oriented at an angle of \(\varphi/2\) relative to the radial direction and appears as a similar distribution.
For \(m=-1\) and \(\varphi=0\), \(\ket{\bm{\mathrm{B}}_{-1}}=[\mathrm{cos}\Theta, -\mathrm{sin}\Theta]^T\) and \(l=-1\). In this case, the polarization toggles between outward and inward radial direction every \(\pi/2\) in azimuth, completing four reversals in a full circle. Since the magnetic polarization in the transitional region forms a vortex distribution during each alternation, four vortex configurations of polarization emerge overall. 
When \(\varphi\neq0\), the magnetic field direction at each position is rotated by \(\varphi/2\), but the overall magnetic field distribution retains a similar configuration. With multiple vortex regions resembling skyrmion in-plane moments, these field configurations enable multi-skyrmion encoding in magnets.

Particularly, when \(\theta=0\) and \(\pi\), it corresponds to right-hand polarized (RCP) and left-hand polarized (LCP) beams on the poles. Notably, their magnetic fields are topologically similar to equatorial beams. Though locally rotating, each instantaneous distribution matches a specific equatorial case \cite{zhang2026}.

When \(\theta\neq0,\pi/2\) or \(\pi\), the HOPS beam corresponds to elliptically polarized beam on the spherical surface (excluding the equator and the poles). Still, they remain topologically similar to equatorial beams. Elliptically polarized beams exhibit local magnetic field rotation (with time-varying amplitude), yet maintain an overall distribution akin to circularly polarized beams. 

Figure. \ref{Fig1}(c) schematically illustrates the light-magnet interaction. We numerically investigate the influence of the magnetic fields of the HOPS structured beams with order \(m=\pm1\) on two distinct types of DMI magnets that respectively host Bloch-type skyrmions and antiskyrmions. We consider a ferromagnetic system comprising 201×201 lattice sites under periodic boundary conditions. The initial magnetic moment at each lattice site is set to (0, 0, 1). The Hamiltonian is given by
\begin{equation}
\begin{aligned}
H = -J\sum\limits_{\bm{\mathrm{r}}}\bm{\mathrm{m}}_{\bm{\mathrm{r}}} \cdot (\bm{\mathrm{m}}_{\bm{\mathrm{r}}+a\bm{\mathrm{e}}_x} + \bm{\mathrm{m}}_{\bm{\mathrm{r}}+a\bm{\mathrm{e}}_y}) \\
+ D\sum\limits_{\bm{\mathrm{r}}, i}\bm{\mathrm{e}}_i \cdot (\bm{\mathrm{m}}_{\bm{\mathrm{r}}} \times \bm{\mathrm{m}}_{\bm{\mathrm{r}}+a\bm{\mathrm{e}}_i}) \\
- H_z\sum\limits_{\bm{\mathrm{r}}}\bm{\mathrm{m}}_{z} - \sum\limits_{\bm{\mathrm{r}}}\bm{\mathrm{B}}(\bm{\mathrm{r}}, t) \cdot \bm{\mathrm{m}}_{\bm{\mathrm{r}}}. 
\end{aligned} \tag{5}
\end{equation}
Here, \(J\), \(D\) and \(H_{z}\) represent the exchange constant, the DMI constant and the constant external magnetic field applied along the \(z\) axis, \(a\) the lattice constant, \(\bm{\mathrm{e}}_i\) the unit vector oriented along either the \(x\) or \(y\) axis. Two types of DMI Hamiltonian that generate Bloch-type and antiskyrmions are respectively given by \(H_{DMI}=D\sum\limits_{\bm{\mathrm{r}}, i}\bm{\mathrm{e}}_x \cdot (\bm{\mathrm{m}}_{\bm{\mathrm{r}}} \times \bm{\mathrm{m}}_{\bm{\mathrm{r}}+a\bm{\mathrm{e}}_x})+D\sum\limits_{\bm{\mathrm{r}}, i}\bm{\mathrm{e}}_y \cdot (\bm{\mathrm{m}}_{\bm{\mathrm{r}}} \times \bm{\mathrm{m}}_{\bm{\mathrm{r}}+a\bm{\mathrm{e}}_y})\) and \(D\sum\limits_{\bm{\mathrm{r}}, i}\bm{\mathrm{e}}_x \cdot (\bm{\mathrm{m}}_{\bm{\mathrm{r}}} \times \bm{\mathrm{m}}_{\bm{\mathrm{r}}+a\bm{\mathrm{e}}_y})+D\sum\limits_{\bm{\mathrm{r}}, i}\bm{\mathrm{e}}_y \cdot (\bm{\mathrm{m}}_{\bm{\mathrm{r}}} \times \bm{\mathrm{m}}_{\bm{\mathrm{r}}+a\bm{\mathrm{e}}_x})\).
The spin dynamics is governed by the LLG equation:
\begin{equation}
 \dfrac{d\bm{\mathrm{m}}_{\bm{\mathrm{r}}}}{dt}  = -\gamma \bm{\mathrm{m}}_{\bm{\mathrm{r}}} \times \bm{\mathrm{H}}_{\mathrm{eff}}+\alpha \bm{\mathrm{m}}_{\bm{\mathrm{r}}} \times \dfrac{d\bm{\mathrm{m}}_{\bm{\mathrm{r}}}}{dt} , \tag{6}
\end{equation}
where \(\gamma\) denotes the gyromagnetic ratio, \(\alpha\) the Gilbert damping coefficient, \(\bm{\mathrm{m}}_{\bm{\mathrm{r}}}\) the unit vector of magnetic moment, and \(\bm{\mathrm{H}}_{\mathrm{eff}}=-\nabla _{\bm{\mathrm{m}}_{\bm{\mathrm{r}}}}(H/J)\) the effective magnetic field. Equation (6) is numerically integrated using the fourth-order Runge–Kutta method with a time step of \(\Delta t\).

\begin{figure}
\includegraphics[width=1\linewidth]{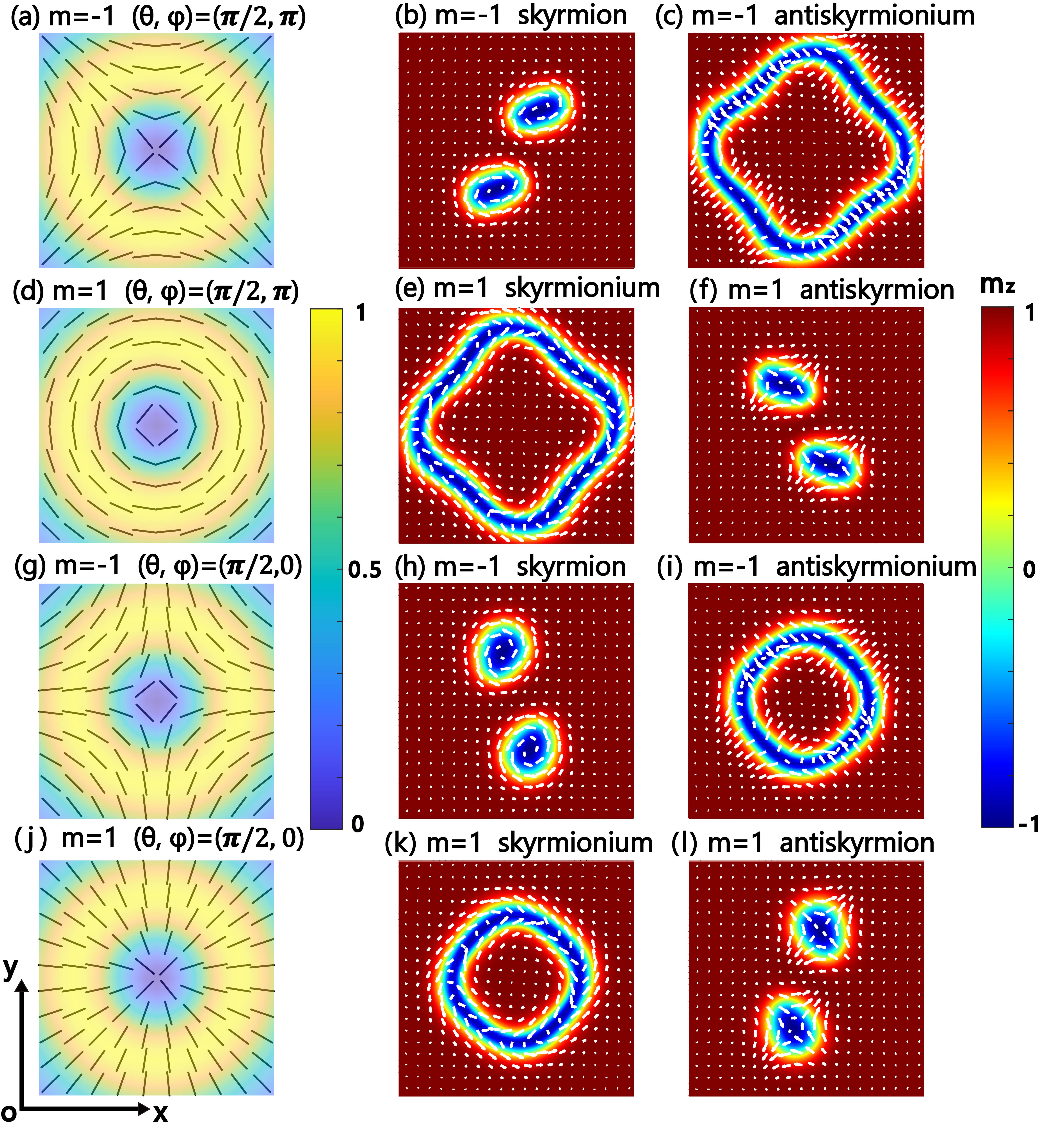}
\caption{\label{Fig2}Magnetization dynamics of the magnets under the influence of beams at \(\theta=\pi/2\) on the HOPS with \(m=\pm1\). Snapshots of spin states for (a), (d), (g), (j) the magnetic fields in magnet possessing DMI that stabilizes (b), (e), (h), (k) Bloch skyrmions and (c), (f), (i), (l) antiskyrmions.
}
\end{figure}

In our simulation, we define one unit of time as \(\hbar/J\), corresponding to 0.505 ps for \(J = 1.3\) meV and one unit of magnetic field density to 22.36 T. Hereafter, we set \(\hbar=J=1\) and nondimensionalize the other quantities. The following parameter values are used in the simulation: \(\Delta t=0.001\), \(\gamma=1\), \(\alpha=0.1\), \(\omega=0.285\) corresponding to about \(0.09\) THz and lattice constant \(a=1\). 

\(Results\)—We first consider the magnetic fields of \(\pi\)-vector beams on the equator (\(\theta=\pi/2\)). Specifically, we study the magnetic response in two ferromagnetic systems that respectively stabilize Bloch skyrmions and antiskyrmions via DMI, under the magnetic fields of two representative \(\pi\)-vector beams: the \(\pi\)-radial beam (\(\varphi=0\)) and the \(\pi\)-azimuthal beam (\(\varphi=\pi\)). We set \(W = 10a\), \(t_{0} = 7.72\), \(\sigma = 2.57\), \(D = 0.5\), \(B_0=0.46\), and \(H_{z} = 0.115\). In Fig. \ref{Fig2}, we show the simulation results at \(t = 129\).

Fig. \ref{Fig2}(a) and (g) show the magnetic field and intensity distributions of the \(\pi\)-azimuthal vector beam at coordinates (\(\pi/2\),\(\pi\)) and (\(\pi/2\),0) on the HOPS with \(m=-1\), respectively. As shown before, these two magnetic field distributions exhibit similar configurations that can be interpreted as four vortex configurations. They can be applied to a magnet possessing DMI that stabilizes Bloch skyrmions, in both cases, inducing two skyrmions, as shown in Fig. \ref{Fig2}(b) and (h), respectively; while a higher amplitude would yield four skyrmions due to the chirality interplay of light and magnetism \cite{zhang2026}. Alternatively, the similar configurations can be viewed as a hollow core structure resembling the internal texture of an antiskyrmion (saddle surface), which leads to the formation of a single antiskyrmionium in a magnet possessing DMI that stabilizes antiskyrmions, as shown in Fig. \ref{Fig2}(c) and (i), respectively.

Fig. \ref{Fig2}(d) and (j) show the magnetic field and intensity distributions of the magnetic field for the \(\pi\)-azimuthal vector beam at coordinates (\(\pi/2\),\(\pi\)) and (\(\pi/2\),0) on the HOPS with \(m=1\), respectively. The corresponding magnetic field configuration in Fig. \ref{Fig2}(d) can be interpreted as a single hollow vortex. This magnetic field will induce a single skyrmionium in a magnet possessing DMI that stabilizes Bloch skyrmions, as shown in Fig. \ref{Fig2}(e). Alternatively, the same field can lead to the formation of two antiskyrmions in a magnet with antiskyrmion-stabilizing DMI, as shown in Fig. \ref{Fig2}(f), while a higher amplitude would yield four antiskyrmions. Owing to the identical winding number, the corresponding magnetic field distribution in Fig. \ref{Fig2}(j) excites similar phenomena in the magnet, as shown in Figs. \ref{Fig2}(k) and (l). 

By comparing Figs. \ref{Fig2}(c) and (i), we observe that the antiskyrmionium generated in Fig. \ref{Fig2}(c) is considerably larger than that in Fig. \ref{Fig2}(i), indicating a stronger magnetic influence from the light in Fig. \ref{Fig2}(a) on the magnet compared to that from Fig. \ref{Fig2}(g). Similarly, the comparison between Figs. \ref{Fig2}(e) and (k) also exhibits a similar phenomenon. This is due to the varying degree of matching between the magnetic polarization distribution of equatorial light (determined by \(\varphi\)) and the magnetization distribution of the in-plane components of the encoded (anti)skyrmions (determined by the helicity number \cite{zhang2020skyrmion,PhysRevB.105.014444}). When \(\varphi=0\), the magnetic polarization distribution of the light is radial, as shown in Figs. \ref{Fig2}(g) and (j). When \(\varphi=\pi\), the magnetic polarization distribution of the light is rotated by \(\pi/2\) compared to that at \(\varphi=0\), resulting in an azimuthal distribution, as shown in Figs. \ref{Fig2}(a) and (d). Constrained by the form of the DMI, this study only encodes (anti)skyrmions with a helicity number of \(\pi/2\). The magnetization distribution of this type of (anti)skyrmion matches the magnetic polarization distribution of light at \(\varphi=\pi\). Therefore, the influence of the magnetic polarization distribution of light at \(\varphi=\pi\) is stronger. 

\begin{figure}
\includegraphics[width=1\linewidth]{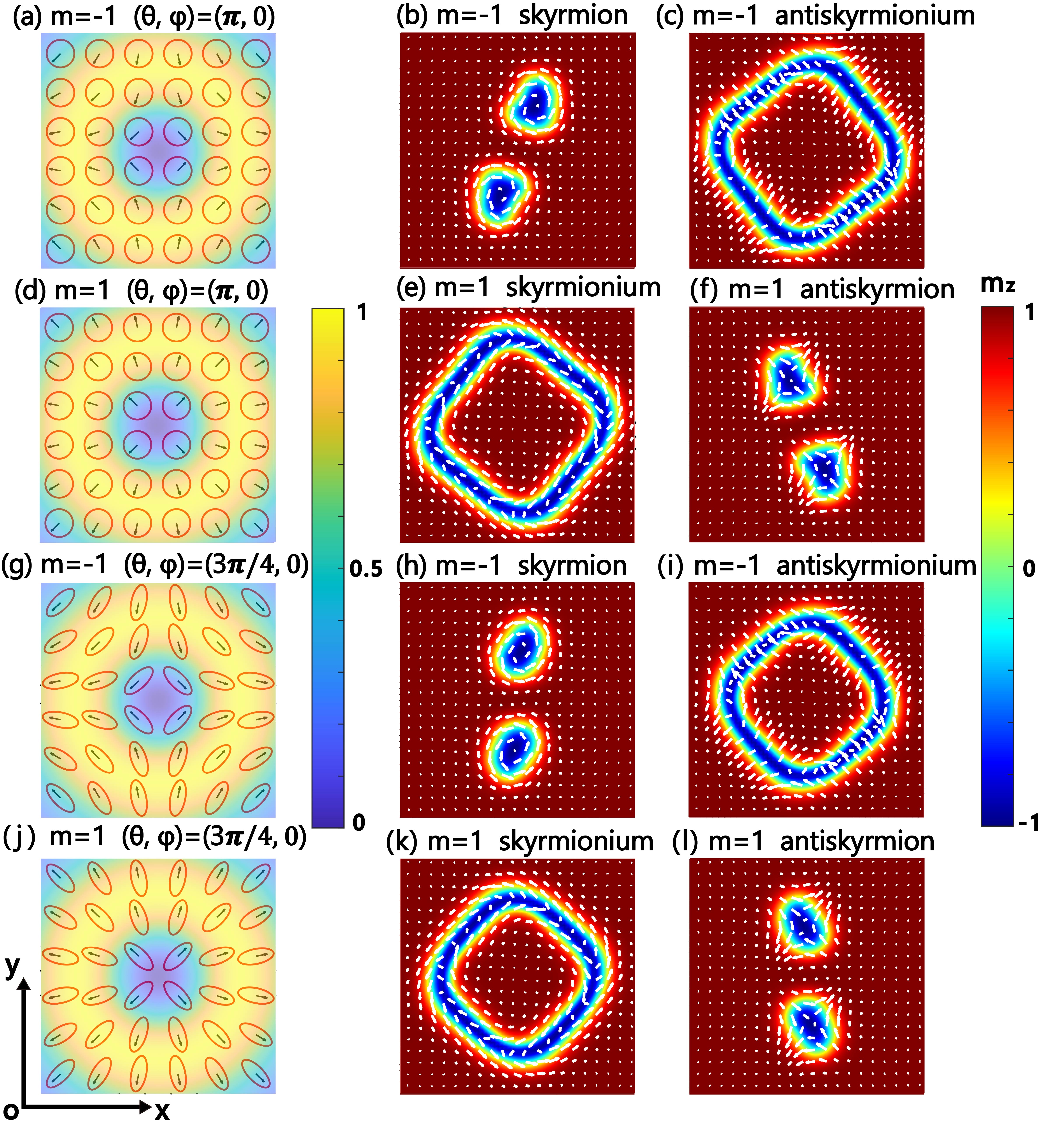}
\caption{\label{Fig3}Magnetization dynamics of the magnets under the influence of beams at \(\theta=\pi\) and \(3\pi/4\) on the HOPS with \(m=\pm1\). Snapshots of spin states for (a), (d), (g), (j) the magnetic fields in magnet possessing DMI that stabilizes (b), (e), (h), (k) Bloch skyrmions and (c), (f), (i), (l) antiskyrmions.
}
\end{figure}

Next, we consider the magnetic fields of  circularly and elliptically (\(\theta\neq\pi/2\)) polarized beams. As established previously, these beams share topological properties with equatorial beams and thus induce similar topological textures. In Fig. \ref{Fig3}, we show the result of the magnetic effects induced by the optical fields at coordinates (\(\pi\),0) and (\(3\pi/4\),0) on the HOPSs for \(m=\pm1\) at \(t = 129\) under the same conditions as above.

Figure \ref{Fig3}(a) and (g) show the magnetic field and intensity distributions of the beams at coordinates (\(\pi\),0) and (\(3\pi/4\),0) on the HOPS with \(m=-1\), respectively. They can induce two skyrmions, as shown in Fig. \ref{Fig3}(b) and (h), and they can also induce a single antiskyrmionium, as shown in Fig. \ref{Fig3}(c) and (i).

Figure \ref{Fig3}(d) and (j) show the magnetic field and intensity distributions of the beams at coordinates (\(\pi\),0) and (\(3\pi/4\),0) on the HOPS with \(m=1\), respectively. They can encode a single skyrmionium, as shown in Fig. \ref{Fig3}(e) and (k) and also encode two antiskyrmions, as shown in Fig. \ref{Fig3}(f) and (l), while a higher amplitude would yield four antiskyrmions.

Based on the results from Figs. \ref{Fig2} and \ref{Fig3}, we conclude that a HOPS of order m consists of lights with an identical winding number, which is given by \(l=m\). The corresponding magnetic fields induce topological textures in the magnet, which may exist as a single entity or multiple entities, with a total topological skyrmion number ranging from 
\begin{equation}
Q_{\mathrm{t}}=Q\cdot|Q_{\mathrm{v}}\cdot m-1| \sim  2Q\cdot|Q_{\mathrm{v}}\cdot m-1|. \tag{7}
\end{equation}
Here, \(Q\) and \(Q_{\mathrm{v}}\) denote the topological skyrmion number and vorticity number of a single skyrmion \cite{zhang2020skyrmion}. \(Q\) is given by \cite{channa2025signatures}: 
\begin{equation}
 Q=\dfrac{1}{4\pi}\int d^2\bm{\mathrm{r}}[\bm{\mathrm{m}}_{\bm{\mathrm{r}}} \cdot (\partial_{x}\bm{\mathrm{m}}_{\bm{\mathrm{r}}}\times\partial_{y}\bm{\mathrm{m}}_{\bm{\mathrm{r}}})]. \tag{8}
\end{equation}
In our simulations, a skyrmion possesses \(Q=-1\) and \(Q_{\mathrm{v}}=+1\), whereas an antiskyrmion possesses \(Q=+1\) and \(Q_{\mathrm{v}}=-1\). In a previous study, we have summarized the relationship between the total angular momentum \(\mathcal{J}\) of circularly polarized vortex light and  \(Q_{\mathrm{t}}\) of skyrmions induced in magnets, that is, from \(Q_{\mathrm{t}}=Q\cdot|\mathcal{J}|\) to \(Q_{\mathrm{t}}=2Q\cdot|\mathcal{J}|\) \cite{zhang2026}. This relationship holds for circularly polarized light, but itself is limited in that it cannot be extended to other polarization states on the HOPS.

To date, an all-optical method to encode skyrmion bags is not yet possible. Strikingly, our all-optical control method enables the encoding of two types of skyrmion bags: the \(S_1(Q_{{\mathrm{bag}}}+1)\) type, with an outer skyrmion enclosing inner skyrmions \cite{bo2023controllable} and the \(S_2(|Q_{\mathrm{bag}}|)\) type, with an outer closed domain enclosing inner skyrmions \cite{kern2025controlled}. \(Q_{\mathrm{bag}}\) represents the topological skyrmion number of the skyrmion bag.

\begin{figure}
\includegraphics[width=1\linewidth]{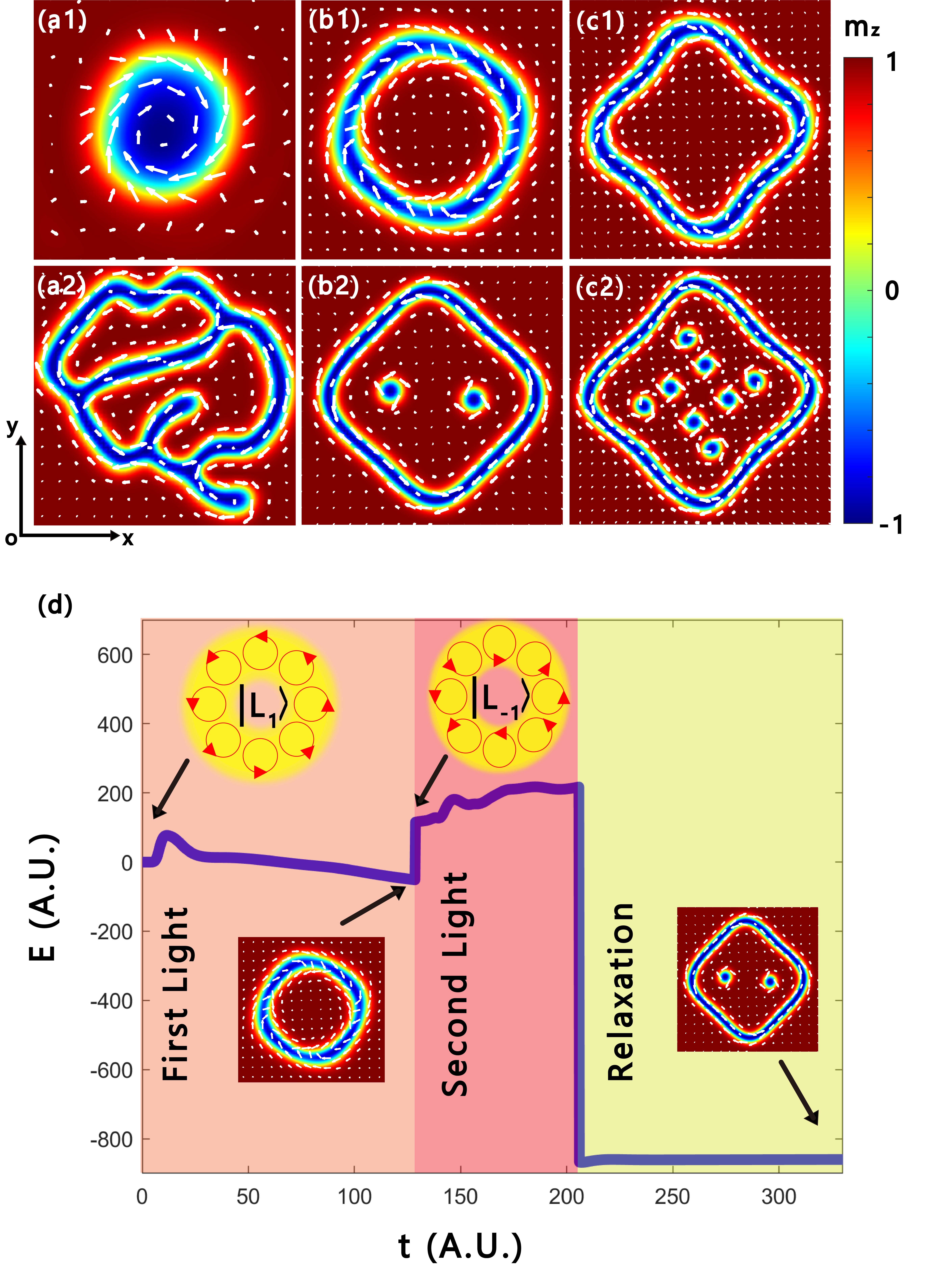}
\caption{\label{Fig4}Magnetization dynamics of magnet possessing DMI that stabilizes Bloch skyrmions under the influence of two kinds of HOPS beams. Snapshots of spin states in the magnet (a1) at \(t=78\) and (a2) in the stable state, which is skyrmion bag \(S_1(2)\); Snapshots of spin states (b1) at \(t=129\) and (b2) in the stable state, which is skyrmion bag \(S_2(2)\); Snapshots of spin states (c1) at \(t=208\) and (c2) in the stable state, which is skyrmion bag \(S_2(8)\); (d) Energy-versus-time curve for the topological texture formation process shown in (b2). 
}
\end{figure}

Building on the above encoding principle, we propose a sequential excitation strategy: sequentially applying two HOPS beams to the magnet a skyrmion bag can be designed and encoded. In Fig. \ref{Fig4}, we show the simulation about the encoding of skyrmion bags.
Fig. \ref{Fig4}(a1) shows the state of a single skyrmion generated in the magnet with \(D = 0.615\) and \(H_{z} = 0.06\) by a RCP beam with \(W=10a\), \(B_0=0.58\), \(t_{0} = 7.72\), \(\sigma = 2.57\) corresponding to the point (0, 0) on a HOPS with \(m=0\), captured at \(t=78\). Subsequently, we use a \(\pi\)-radial vector beam with \(W=5a\), \(B_0=0.19\), \(t_0=10.29\), \(\sigma=5.14\) (corresponding to the point (\(\pi/2\),0) on the HOPS with \(m=-1\)) into the pre-existing skyrmion. Since the skyrmion bag has negative formation energy, it may subsequently undergoes a stripe phase evolution
to lower its energy, eventually transforming into stripe domains \cite{wang2021stripe}. Thus, we increase the external magnetic
field to arrest this energy-driven evolution and stabilize
the configuration. Specifically, at \(t=155\), we increase the external magnetic field to \(H_z=0.192\) and allow the system to relax. Finally, a stabilized skyrmion bag \(S_1(2)\) is shown in Fig. \ref{Fig4}(a2).

Further, Fig. \ref{Fig4}(b1) shows the state of a single skyrmionium generated in the magnet with \(D = 0.5\) and \(H_{z} = 0.115\) by the LCP beam with \(W=10a\), \(B_0=0.58\), \(t_{0} = 7.72\), \(\sigma = 2.57\) corresponding to the point (\(\pi\),0) on the HOPS with \(m=1\), captured at \(t=129\). We then irradiate this pre-existing skyrmionium with a \(\pi\)-radial vector beam (corresponding to the point (\(\pi/2\),0) on the HOPS with \(m=-1\) ) with \(W=10a\), \(B_0=0.27\), \(t_0=10.29\), \(\sigma=5.14\). Next, at \(t=206\), we increase the external magnetic field to \(H_z=0.192\) and allow the system to relax. Finally, stabilized skyrmion bag \(S_2(2)\) is shown in Fig. \ref{Fig4}(b2). We set the initial energy of the system to zero, and the energy evolution throughout the process is shown in Fig. \ref{Fig4}(d), where the energy of the skyrmion bag eventually stabilizes. The energy evolution behaviors of all skyrmion bags in Fig. \ref{Fig4}(a-c) are similar.

Our strategy can also be extended to the case with \(|m|>1\). However, the pre-existing skyrmionium in Fig. \ref{Fig4}(b1) fractures under irradiation by a \(\pi\)-radial vector beam at (\(\pi/2\),0), as the higher OAM increases the beam’s effective area. Specifically, the resulting full coverage fractures the structure by forcing its internal moments to reorient. To prevent this, we first evolve the skyrmionium under the same configuration above until \(t=206\) in Fig. \ref{Fig4}(c1). The skyrmionium exhibits a larger size and enhanced structural stability, making it less susceptible to fracturing upon subsequent optical excitation. The underlying mechanism for this size increase
is its negative formation energy; to approach the energy
minimum, it spontaneously expands and gradually develops
striations, eventually evolving toward a stripe phase \cite{wang2021stripe}. Subsequently, we applied a \(\pi\)-radial vector beam (with \(W=10a\), \(B_0=0.46\), \(t_0=10.29\), \(\sigma=5.14\)) from the point (\(\pi/2\),0) on the HOPS with \(m=-3\). Next, at \(t=283\), we increase the external magnetic field to \(H_z=0.192\) and allow the systems to relax. Finally, a stabilized skyrmion bag \(S_2(8)\) is shown in Fig. \ref{Fig4}(c2). Videos illustrating the encoding of skyrmion bags in Fig. \ref{Fig4} can be found in the Supplemental Material. 

\(Conclusion\)—In summary, we develop a theoretical framework that establishes a topological connection between structured light and magnetic texture. For any polarized light on the HOPS, the structured magnetic fields generated therefrom share similar topological properties and can encode varying numbers of skyrmions and antiskyrmions on two distinct types of DMI magnets that respectively host Bloch-type skyrmions and antiskyrmions. Specifically, when the order is \(m\), topological textures with an absolute \(Q_{\mathrm{t}}\) ranging from \(Q\cdot|Q_{\mathrm{v}}\cdot m-1|\) to \(2Q\cdot|Q_{\mathrm{v}}\cdot m-1|\) can be encoded. Furthermore, by sequentially applying two HOPS structured beams to the magnetic material, skyrmion bags with different \(Q_{\mathrm{bag}}\) can be encoded.

Our all-optical method for encoding topological textures offers broader implications than previous findings that relied solely on circularly polarized light to generate skyrmions \cite{zhang2026}. Here, the scope extends from circular polarization to any polarization on the HOPS, and from only skyrmions to both skyrmions and antiskyrmions. In contrast to the predominantly non-all-optical methods currently available \cite{foster2019two,bo2023controllable,kern2025controlled,yang2024embedded,jiang2025stable,liu2025room}, we provide an all-optical alternative for encoding various types of skyrmion bags in magnetic materials. Our method offers enhanced convenience, efficiency, and flexibility.

\(Acknowledgment\)—S. Lin acknowledges the funding from National Natural Science Foundation of China (Grant No. 12104296) and the startup funding from Great Bay University (No. YJKY220018). Y. Shen acknowledges the funding from Singapore Ministry of Education (MOE) AcRF Tier 1 grants (RG157/23 \& RT11/23), Singapore Agency for Science, Technology and Research (A*STAR) MTC Individual Research Grants (M24N7c0080), and a Nanyang Assistant Professorship Start Up grant. Z. Nie acknowledges the funding from National Natural Science Foundation of China (Grants Nos. 12574333 and 11974258), Innovation Research Fundation of National University of Defense Technology (Grant No. 25-28).

\nocite{*}
\bibliographystyle{unsrtnat} 
\bibliography{panjialai}

\end{document}